\documentclass[11pt,twoside]{article}
\usepackage{asp2010}

\pdfoutput=1

\usepackage{graphicx}

\resetcounters

\begin{document}

\title {Starlight and Sandstorms: \\
        Mass Loss Mechanisms on the AGB}
\author{Susanne H{\"o}fner
\affil{Department of Physics \& Astronomy,
       Uppsala University,
       Sweden }
}

\begin{abstract}
There are strong observational indications that the dense slow winds
of cool luminous AGB stars are driven by radiative pressure on dust
grains which form in the extended atmospheres resulting from
pulsation-induced shocks. For carbon stars, detailed models of
outflows driven by amorphous carbon grains show good agreement with
observations. Some still existing discrepancies may be due to a
simplified treatment of cooling in shocks, drift of the grains
relative to the gas, or effects of giant convection cells or
dust-induced pattern formation. ~For stars with \,C/O $<$ 1, recent
models indicate that absorption by silicate dust is probably
insufficient to drive their winds. A possible alternative is
scattering by Fe-free silicate grains with radii of a few tenths of
a micron. In this scenario one should expect less circumstellar
reddening for M- and S-type AGB stars than for C-stars with
comparable stellar parameters and mass loss rates.

\end{abstract}


\noindent {\em ``It is a truth universally acknowledged, that an
evolved cool giant star must develop a wind. However little known
the parameters of such a star may be on its first entering the red
giant stage, this truth is so well fixed in the minds of stellar
astronomers, that it is considered as the rightful member of some
one or other class of mass-losing long-period variables."}

\noindent
~---~Introduction to a fictitious manuscript entitled `Winds of Cool
Giants: Properties and Prejudices,' adapted from a classical text by
\citet{austen13}


\section{Introduction}

Winds of AGB stars can be studied observationally with a variety of
methods, ranging from classical spectroscopy and photometry through
direct imaging to interferometric techniques. These various methods
are complementary in the sense that they can provide information
about stellar winds on very different spatial and temporal scales,
as well as independent ways of studying a particular effect. From
photometric time series we can deduce crucial information about
stellar pulsation and possibly dust formation; high-resolution
spectra and IR multi-wavelength interferometry allow quantitative
insights into the dynamics and structure of the regions where wind
formation takes place; and imaging of circumstellar envelopes on
global scales holds clues to the mass loss history of individual
stars and their interaction with the surrounding interstellar
medium. The general picture derived from observations is that of
largely spherical, but probably clumpy and time-dependent, outflows
with typical velocities of 5--30 km/s and mass loss rates of about
$10^{-7}$ to $10^{-5}$\,M$_{\odot}$/yr.

A thorough physical understanding of the mass loss phenomenon and
detailed quantitative models of stellar winds are required to
understand both AGB stars in their own right, and their role in the
bigger picture of the cosmic matter cycle. Reliable mass loss
rates, dust production rates, and consistent synthetic spectra are
necessary ingredients for population synthesis, modelling the chemical
evolution of galaxies, and gauging the contribution of AGB stars to
the integrated light and intrinsic reddening of distant galaxies.

\begin{figure}[t]
\begin{center}
\includegraphics
[width=10cm] {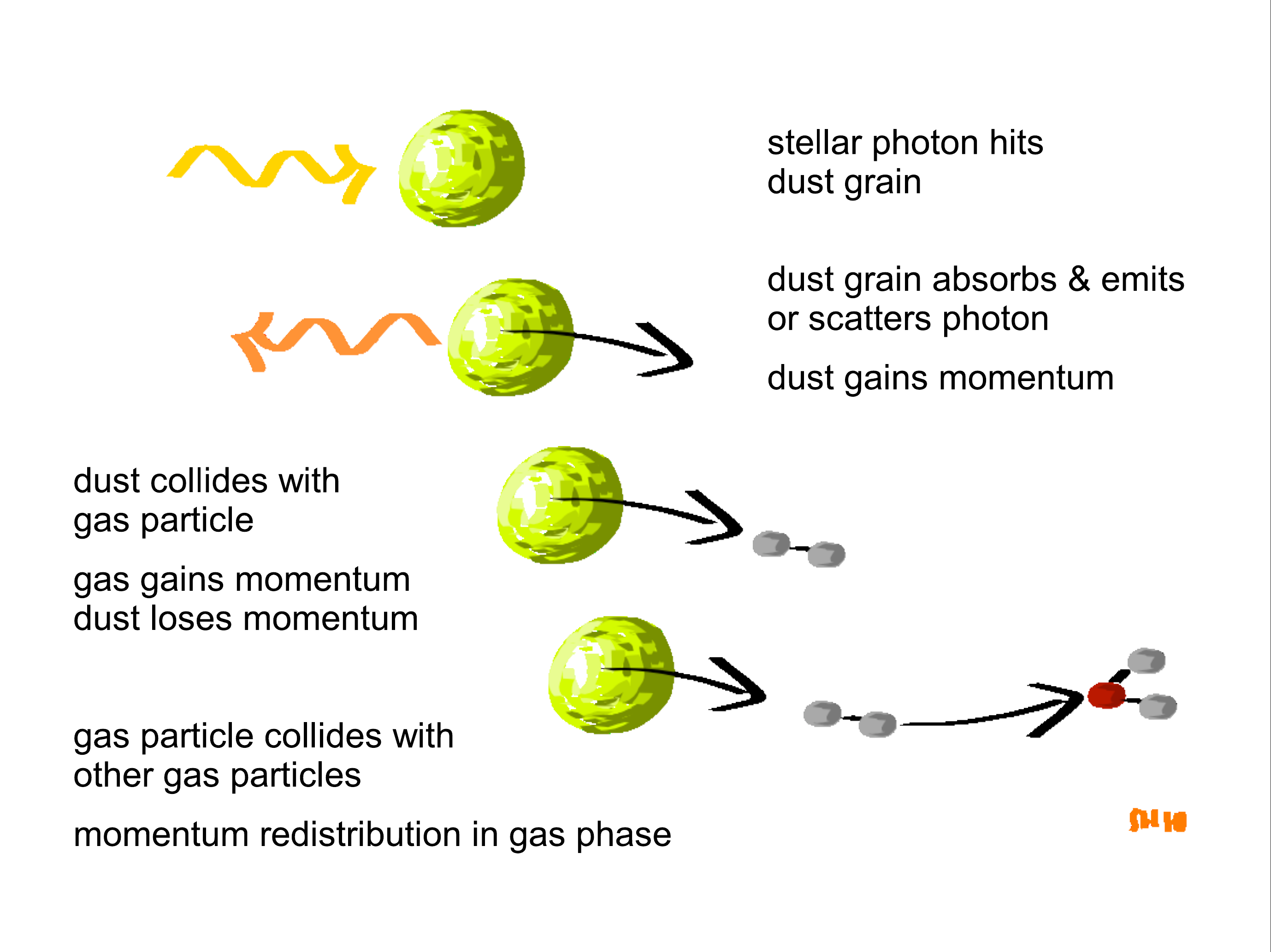}
\end{center}
\caption{Microphysics of dust-driven winds: Dust grains acquire
momentum from stellar photons and transfer it to the surrounding gas
via collisions.} \label{f_drive}
\end{figure}

\section{Basic Scenarios and Wind Mechanisms}\label{s_basics}

A common feature of the most evolved cool giants is the presence of
dust in their outer atmospheres and winds, often inferred from
photometry (circumstellar reddening, IR excess), and sometimes more
specifically from the detection of spectral features characteristic
of particular species of dust grains (see, e.g., Waters, this volume). 
In principle, the grains could just be a by-product of the outflows,
condensing from the cooling gas as it moves away from the stellar
surface. It has, however, been suspected for a long time that
radiative acceleration of dust is an important ingredient of the
wind driving mechanism \citep[see Fig.~\ref{f_drive}; for a historical
overview see, e.g.,][]{ho04}.

The formation and survival of dust particles requires temperatures
below the stability limit of the respective condensate. As the
temperature of a grain will be mostly determined by its interaction
with the radiation field, it cannot exist closer to the star than
the distance where its radiative equilibrium temperature is equal
to the condensation temperature $T_c$ of the grain material.
Assuming a Planckian stellar radiation field, and a power law for
the grain absorption coefficient $\kappa_{\rm abs} \propto
\lambda^{-p}$ in the relevant wavelength range (i.e. around the flux
maximum of the star), the condensation distance $R_c$ can be
estimated by
  $ R_c/R_{\ast} =
    0.5 \left( T_c/T_{\ast} \right)^{- (4+p)/2} $
\citep[see, e.g.,][]{lame99}. For amorphous carbon grains
with $T_c \approx 1500\,$K and $p \approx 1$ we obtain $R_c/R_{\ast}
\approx 2-3$ which compares well with detailed models.

\begin{figure}[t]
\begin{center}
\includegraphics
[width=12cm] {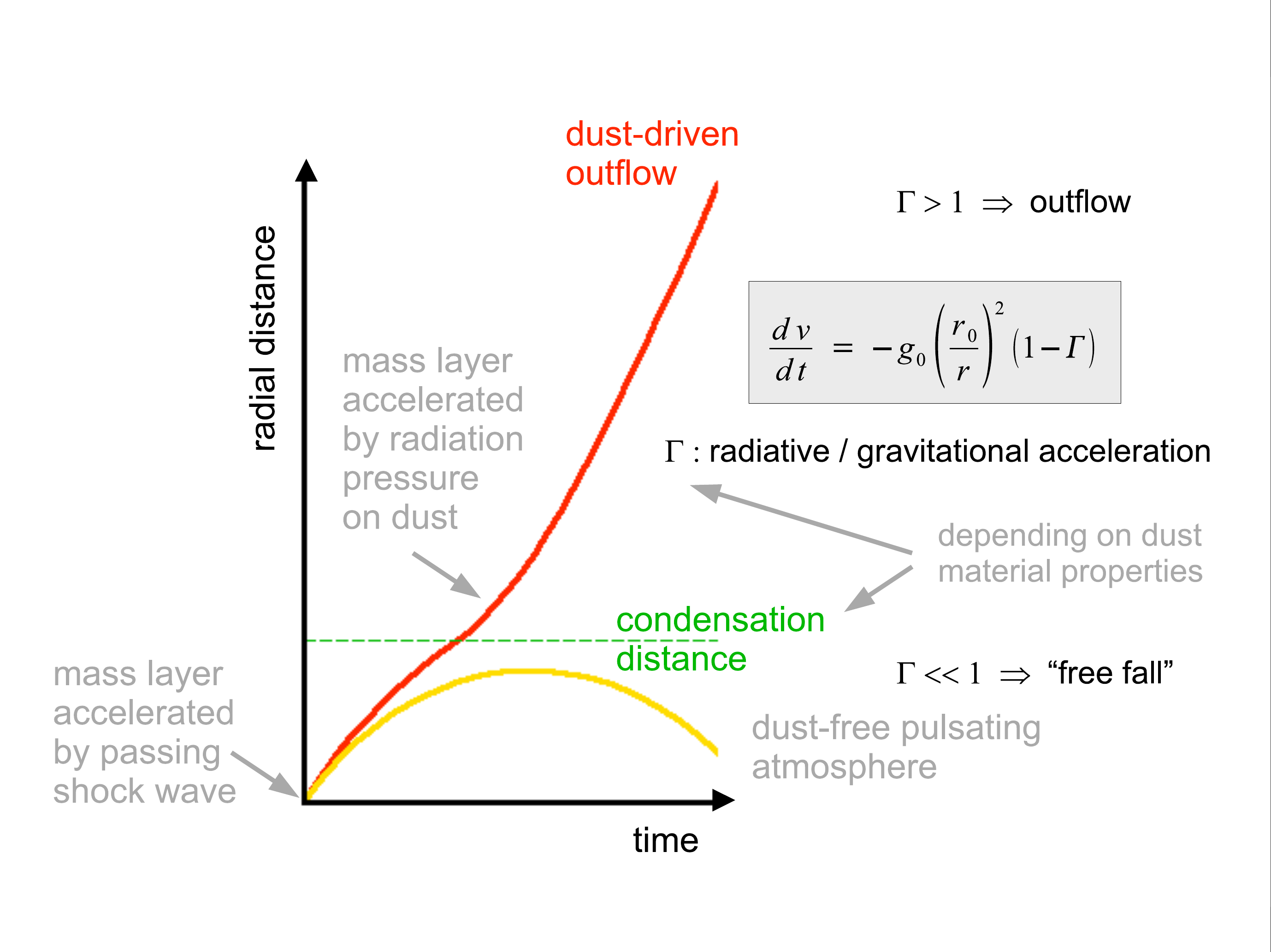}
\end{center}
\caption{Schematic picture of the dynamics of mass layers, based on
a simple toy model taking only gravity and radiative acceleration
into account \citep[see][for details and typical numbers]{hoefner09}.}
\label{f_dyn}
\end{figure}

At this point, the question arises how the dust-free gas can bridge
the gap between the stellar photosphere and the condensation
distance. Some early models, assuming a gradual transition from a
nearly hydrostatic atmosphere to a steady, time-independent outflow,
invoked Alfv{\'e}n waves \citep[e.g.][]{hart80} or
acoustic wave pressure \citep[e.g.][]{pij89} to start or even
drive the winds. Such mechanisms may be important for stars with
parameters that exclude dust-driven winds, but for cool AGB stars
with large-amplitude pulsations the combination of atmospheric
levitation due to shock waves and radiative acceleration of dust
grains seems to be more effective. The shock waves triggered by the
pulsation periodically accelerate the upper atmospheric layers
outwards, intermittently creating dense environments in their wakes.
The layers follow approximately ballistic trajectories which may
take them above the condensation distance if the initial velocity is
high enough. There, dust condensation can occur and radiation
pressure may accelerate the dust-gas mixture outwards (see
Fig.~\ref{f_dyn}). This scenario is supported by spectroscopic and
interferometric observations of extended, dynamical molecular layers
around AGB stars \citep[see, e.g., contributions by Ireland, Wittkowski 
et al., Ruiz-Velasco et al., all this volume;][]{tlsw03a, wein04, sacu10}.

Time-dependent wind models investigating the effects of pulsation and 
radiative acceleration
of dust were \,pioneered \,by \,\citet{w79} \,and \,\citet{b88}, 
\,using a simple parameterized description for the dust opacity, 
followed by studies including time-dependent grain growth for C stars 
\citep[e.g.][]{fgs92, hd97}. Detailed models
based on the `pulsation-enhanced dust-driven wind scenario' have been 
quite successful in reproducing typical mass loss rates and wind 
velocities, as well as photometry and spectra at various resolutions 
\citep[e.g.][and this volume; and Eriksson et al., this volume]{bertre98, 
wljhs00, andersen_etal03, loidl_etal04, nowo10}. While an
earlier generation of models with grey radiative transfer was
sufficient to explain the main characteristics of heavily
dust-enshrouded C-rich AGB stars, it is necessary to combine
frequency-dependent radiative transfer (including gas and dust
opacities) with time-dependent hydrodynamics and non-equilibrium
dust formation, in order to obtain realistic results for objects
with less optically thick envelopes \citep[e.g.][]{hoefner_etal03}.

\begin{figure}[t]
\begin{center}
\includegraphics
[width=12cm] {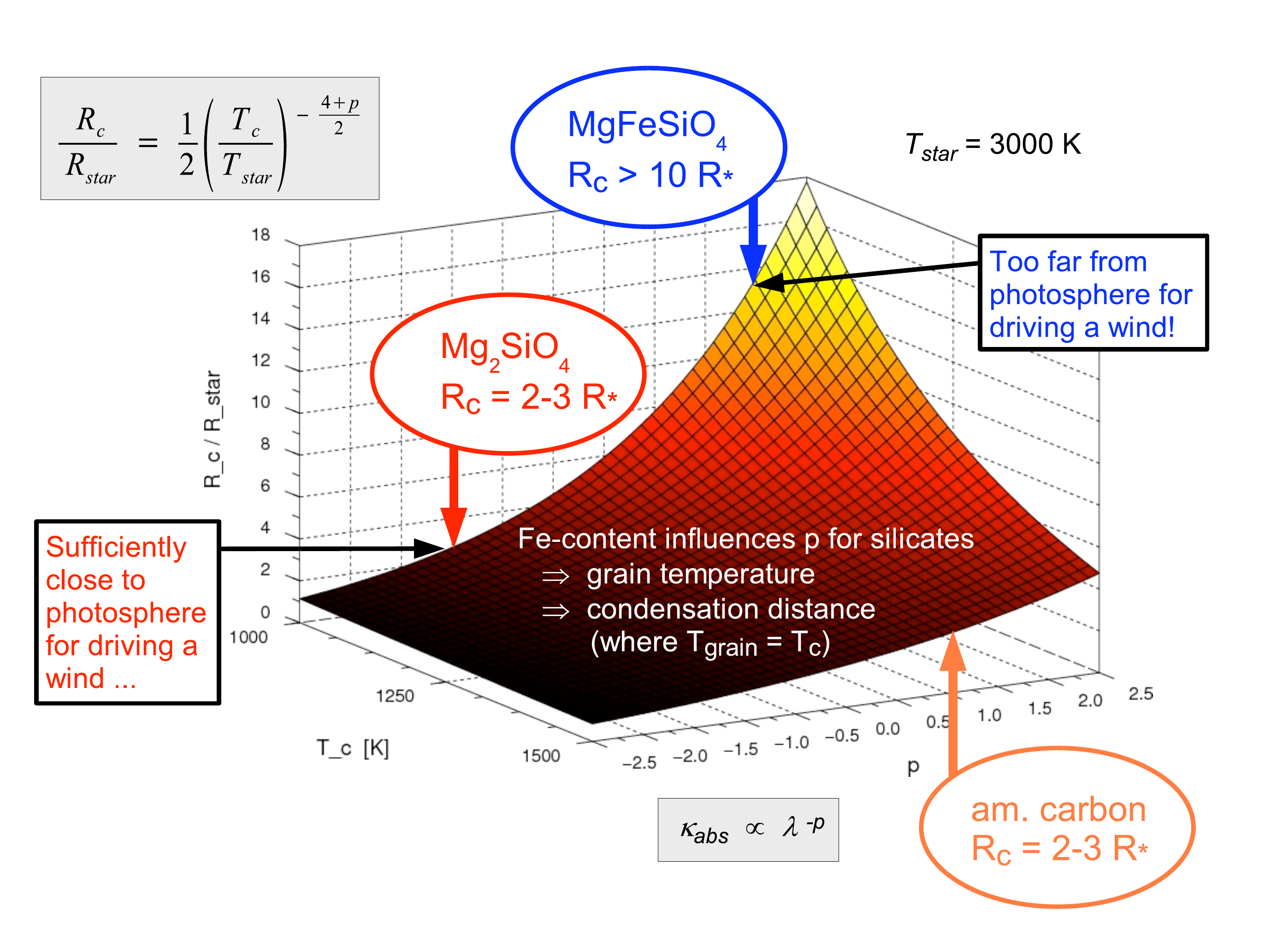}
\end{center}
\caption{Condensation distance $R_c$ (in units of the stellar radius
$R_{\ast}$) as a function of the power law index of the absorption
coefficient $p$ for a range of condensation temperatures $T_c$
(assuming a stellar temperature of $T_{\ast} = 3000\,$K; see text
for details). } \label{f_rc}
\end{figure}

Ironically, it was the introduction of non-grey dynamical models
which led to a crisis regarding the role of dust as a wind driver
for M-type AGB stars. In contrast to C-type objects (where the
excess carbon not bound in CO can condense into amorphous carbon
grains which can drive outflows), the more common AGB stars with C/O
$< 1$ have no abundant chemical elements which can form dust on
their own sufficiently close to the stellar surface. Based on
relative abundances, chemical properties, and thermodynamical
conditions, it is commonly assumed that olivine- and pyroxene-type
Mg--Fe silicates are the main dust species in M-type stars 
\citep[e.g.][Andersen, this volume; Waters, this volume]{GS99, 
ferra01, gail03}. Using detailed non-grey dynamical
models, \citet{woit06b} demonstrated that silicate grains have to be
virtually Fe-free at distances corresponding to the wind
acceleration zone, leading to insufficient radiative pressure due to
low absorption cross sections. The core of the issue can be
understood using the simple estimate for the condensation distance
$R_c$ given above: For silicates, $T_c \approx 1000\,$K and $p$ is
strongly dependent on the Mg/Fe ratio. Considering olivine-type
material, grains with about equal amounts of Fe and Mg lead to $p
\approx 2$ for the absorption coefficient of small grains and,
consequently, to $R_c/R_{\ast} > 10$, whereas iron-free particles
with a corresponding value of $p \approx -1$ can form at typically
$R_c/R_{\ast} \approx 2-3$ (see Fig.~\ref{f_rc}; comparable numbers
hold for pyroxene-type particles).

According to models by \citet{hoefner08a} this problem can be
resolved if conditions in the extended atmosphere allow \,Fe-free
\,silicate grains to grow to \,sizes \,$> 0.1~\mu$m because scattering
will contribute significantly to the radiative pressure for grains
with radii comparable to wavelengths near the stellar flux maximum.
Various observational tests of this scenario are currently being
performed as described in the following section. In this context, it
should be noted that the recent C-star wind model grid by 
\citet{matt10} indicates that the amorphous carbon grains forming in
these stars may also be bigger than previously assumed. In this
case, scattering on grains may contribute to the radiative pressure,
possibly affecting winds close to the thresholds for dust-driven
mass loss (see Mattsson \& H{\"o}fner, this volume).
For typical winds of C stars, however, grain size will not be a
decisive factor, as even small amorphous carbon particles are
efficient absorbers and can easily drive outflows.

\section{A Closer Look: Differences between M- and C-type AGB Stars}

From the early days of stellar spectroscopy, cool giants could be
sorted into two major groups, according to the relative abundances
of C and O in their atmospheres and the resulting distinct
differences in molecular chemistry.\footnote{This dichotomy is
caused by the high bond energy of CO which leads to an almost complete
blocking of the less abundant of the two elements in CO. In C-type
objects (C/O $>$ 1) the excess carbon can form C-bearing molecules
and amorphous carbon grains, while M-type stars (C/O $<$ 1) feature
O-bearing species in their molecular chemistry and dust.} 
With the advent of space-based
IR surveys, however, it became evident that this chemical difference
even leads to profound effects in the circumstellar envelopes. Using
photometry (e.g. the classical IRAS 2-colour diagram), AGB stars can
be sorted both according to current mass loss rates and mass loss
history (detached shells), as well as by the C/O ratio. C-type AGB
stars tend to show much more pronounced circumstellar reddening at
visual and NIR wavelengths than M-type objects with comparable
stellar parameters. In view of the clear spectroscopic and
photometric distinctions between M- and C-type objects, it is
interesting to note that the average wind properties of M-, S-, and
C-type AGB stars are very similar \citep[e.g.][]{ramstedt}.

For cool luminous carbon stars there is strong evidence that their
winds are driven by radiation pressure on amorphous carbon grains.
Models based on the pulsation-enhanced dust-driven wind scenario
(discussed in the previous section) have been tested against a range
of observations and are now used for the interpretation of
observational data of individual stars, and applied to stellar
evolution \citep[e.g.][]{kps03}.

For M-type AGB stars, in the light of the detailed models by 
\citet{woit06b} it seems unlikely that winds can be driven by radiative
pressure on Fe-bearing silicate grains, in contrast to previous
assumptions. As discussed in Sect.~\ref{s_basics}, the inclusion of
Fe in olivine- and pyroxene-type particles\footnote{The presence of
such dust particles can be inferred both from observations of dust
features in the IR (e.g.\ Waters, this volume), and from gas
kinetic arguments \citep[e.g.][]{gail03}. Notably, under chemical
equilibrium conditions the Mg-rich end members of the olivine and
pyroxene sequences should be dominant.} leads to high grain
temperatures and, consequently, condensation distances well beyond
the wind acceleration zone (see Fig.~\ref{f_rc}). Fe-free olivine-
and pyroxene-type grains which can form sufficiently close to the
stellar photosphere, on the other hand, have low absorption cross
sections in the critical wavelength range near the stellar flux
maximum (around $1\,{\mu}$m) which are insufficient for driving an
outflow.

A possible solution of this problem is scattering: If conditions in
the extended atmospheres allow such grains to grow into the size
range of about $0.1 - 1\,\mu$m, scattering becomes dominant over
absorption by several orders of magnitude, opening up the
possibility of stellar winds driven by scattering on Fe-free grains.
Detailed frequency-dependent dynamical models by \citet{hoefner08a}
show that forsterite grains may well reach this critical size range,
leading to outflows with combinations of mass loss rates and wind
velocities that compare well with observations 
\citep[see Fig.~5 in ][]{hoefner09}. First tests of synthetic spectra 
and photometric
colours resulting from these models show good agreement with
observations (see Fig.1 in Bladh et al., this volume), and an
ongoing study of self-consistent M-type wind models with
parameterized dust opacities should allow us to draw further
conclusions about what relative levels of true absorption and
scattering by dust are compatible with observed SEDs (Bladh et al.,
in prep.). Furthermore, new spectro-interferometric data probing the
outer atmosphere and inner wind region of RT Vir shows a clear
transition from a purely molecular to a dusty regime which may give
us direct indications of condensation distances and grain types (see
Olofsson et al., this volume).

If scattering on Fe-free silicate grains turns out to be a major
driving mechanism in dusty M-type (and maybe S-type) AGB stars,
the levels of true absorption by dust (and, consequently, circumstellar
reddening), as well as the dynamical response to dust formation
(the resulting acceleration) may be quite different from that in C-type
AGB stars with similar stellar parameters. These effects might be
related to, for example, recently discovered qualitative differences 
in the phase-to-phase and cycle-to-cycle variations (or lack thereof) 
in IR interferometric data (see Karovicova et al., and Ohnaka, this 
volume).

\section{Open Questions, Conclusions and Outlook}

Current dust-driven wind models focus on the chemical composition
and optical properties of dust grains while often using the
simplifying assumptions of complete momentum coupling and position
coupling for the all-important interaction of dust and gas. In
reality, the grains will move faster than the gas, and this drift
can affect grain growth and wind dynamics \citep[e.g.][]{cssh3}, 
as well as the large-scale structure of the CSE \citep{sid01}. 
Another physical feature of present models which may need
further investigation is strong radiative cooling behind shocks due
to high molecular opacities, corresponding to chemical equilibrium
and LTE conditions. Less efficient cooling could open the
possibility of `pulsation-driven dust-enhanced winds' as discussed
by \citet{will00}. Furthermore, considering the crucial role of
atmospheric dynamics for the wind mechanism, both self-excited
pulsation models (e.g.\ Wood \& Arnett, this volume), and a detailed
analysis of time-dependent phenomena in atmospheres and winds (e.g.\
Dreyer et al., this volume) should be high on the agenda of
modellers.

The `pulsation-enhanced dust-driven wind scenario', discussed in
some detail in this review, seems to be the most promising
explanation for the high mass loss rates of the most evolved cool,
luminous AGB stars, but there are natural limits for this mechanism.
High stellar temperatures lead to large condensation distances, and
low luminosity-to-mass ratios may require more radiative
acceleration than can be provided by grain radiative cross sections.
For low metallicities, the availability of condensable material may
be another issue. Here, again, significant differences between C-
and M-type stars should exist, since carbon stars produce the main
dust-forming element by nucleosynthesis, whereas M-type objects
basically have to rely on the supply of potentially dust-forming
elements with which they started out. Therefore, mass loss should
show a much stronger dependence on metallicity for M-type AGB stars
than for their C-rich counterparts \citep[e.g.][]{2008A&A...484L...5M} 
which seems to fit with observed trends (e.g. Lagadec et al., Sloan 
et al., this volume).

Focusing on the microphysics of the driving mechanism, it is easy to
overlook that observations show evidence of deviations from
homogeneous, spherically symmetric outflows at various scales, e.g.
clumpy structures, large-scale asymmetries of the CSE, and
structures possibly due to focusing of wind material by close
binaries (Kim \& Taam, Mohamed \& Podsiadlowski, this volume). Whether
any of these phenomena are directly related to or give constraints
on the mass loss mechanisms is unclear at present. 3D
`star-in-a-box' models of stellar convection and its effects on dust
formation \citep{frey08} or studies of dust-induced
pattern formation in CSEs \citep{woit06a} are few and far between due
to the considerable numerical effort involved and various technical
difficulties.

Last but not least it should be mentioned that indications of
stellar winds are seen for AGB stars falling outside the parameter
region where present models predict dust-driven mass loss 
\citep[e.g.][]{wach08, matt10}, challenging our current
understanding of mass loss mechanisms. Where both dust and shock
waves fail as wind drivers, magnetic fields and Alfv{\'e}n waves
might play a role in accelerating the observed outflows 
\citep[see, e.g.,][]{hart80, aira10, vlemm10}.

Having started this review with a fictitious quotation, it may be
appropriate to end it with a real one. \citet{prat93} claims that
``Of all the forces in the universe, the hardest to overcome is the
force of habit. Gravity is easy-peasy by comparison." Translating
this to the present context, we may conclude that AGB stars clearly
manage to drive winds, but gaining a comprehensive, quantitative
understanding of the relevant mechanisms may require thinking
outside the current computational box.

\bibliography{hoefner}

\end{document}